# Development of Pipetting Devices to Separate Protein Complexes

Christopher M. Altenderfer, Department of Mechanical Engineering, Temple University

Frank N. Chang, Department of Biology, Temple University

Parsaoran Hutapea, Professor (Corresponding Author)

Composite Laboratory, Department of Mechanical Engineering, Temple University

1947 N 12$^{th}$ Street, Philadelphia, PA 19027

hutapea@temple.edu, Phone: (215) 204-7805, Fax: (215) 204-4956

The objective of this project is to develop an automated device used to spot protein samples on a hydrophobic membrane to be used for the patented electrophoresis method developed by Chang and Yonan in 2008 [1]. This novel method performs electrophoresis directly on hydrophobic blot membranes as opposed to the previous popular methods such as the 2-D polyacrylamide gel method [2, 3]. This new electrophoresis method significantly reduces the processing time to about 40 minutes, as opposed to the 2-D PAGE method which can take one or two days. Special methods to spot samples on hydrophobic blot membranes were established for successful separation of protein and protein complexes. These spotting methods are used in conjunction with the patented electrophoresis method [1] for effective separation. The automated device effectively replicates these special methods that are repeatable and user independent. The goal of the automated device is to effectively spot and separate protein complexes in addition to single proteins under non-denaturing conditions, thus eliminating the variability of the manual spotting method.

The two automated methods proposed to spot the proteins on the surface use a pneumatic air slide in which the pipette moves linearly downward to produce a spotting method that is repeatable. Separation of the proteins using the patented electrophoresis method [1], is highly dependent on the method of spotting. Even manual users following specific procedures, find difficulty in separating the proteins well along the hydrophobic membrane surface after spotting. It is for this reason; a device in which variability is minimized has been created. Through subsequent experiments, it has been shown that our device has the ability to produce a constant result where separation of protein has occurred.

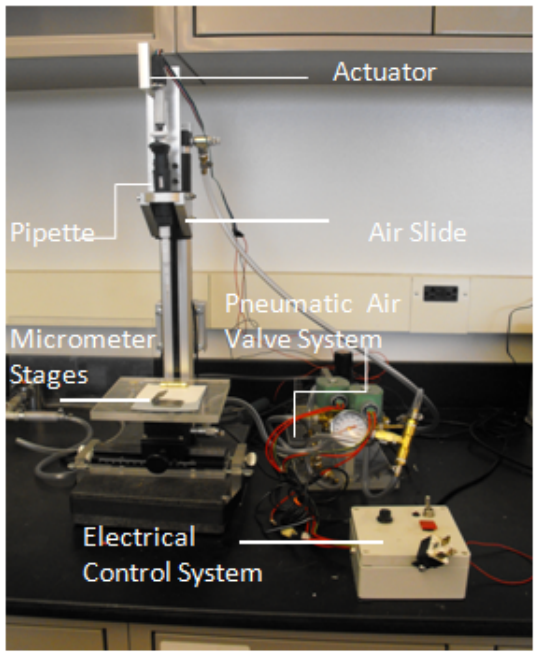 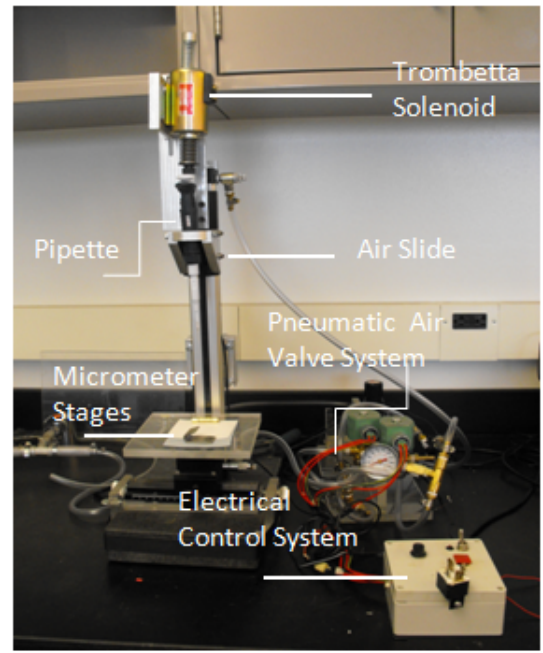

Figure 1. Device System

Figure 1 illustrates the two proposed methods of spotting proteins on the hydrophobic surface. Each method is believed to have unique advantages over the other. Observations of successful manual testing indicate force application is most critical in obtaining good electrophoresis results.

To the left, the touch-and-lift method uses an actuator that preloads the pipette tip such that 70-80 percent of the sample hangs from the tip. As the air slide moves linearly downward, spotting occurs when the sample touches the membrane where capillary action draws the sample from the tip. This touch-and-lift method helps to minimize contact area as the upward motion results in a "Hershey Kiss" like sample. It is through this minimization of contact area that proteins are able to be stacked on the membrane surface and thus prevent binding of the membrane to the surface. Binding of the proteins prevents proteins from effectively separating along the membrane surface.

To the right, the plunger method uses a simple starter solenoid in which high force compresses the pipette when tip to membrane distance is approximately 1-2mm during the downward stroke. This method removes the entire sample from the tip with high force. It is believed that high force, especially near the end of the stroke, is advantageous in creating the necessary conditions for separation. This conclusion is based on manual and automated results in which a high force at the end of the stroke seems to prevent binding from occurring.

Experiments use a Pipetman P2 Pipette and a sample size of 0.5 uL. The solution is 50 percent water and 50 percent ε-caprolactone. Caprolactone is a solvent and acts as an appropriate wetting agent as described in *US Patent No. 7,326,326* [1]. The device system uses a pneumatic air slide in which motion and displacement is controlled using two sensors. Activation of the air slide results in one downward and upward motion in which solenoid or actuator activation occurs automatically. Tip to membrane distance is controlled macroscopically by sensor position and microscopically by micrometer stages. The hydrophobic Polyvinylidene Fluoride (PVDF) membrane is placed on the spotting surface and held stationary as spotting occurs. After spotting occurs, the membrane is immediately taken over to the lab to run electrophoresis. The membrane is first submerged in a low conductivity buffer solution in which the pH is adjustable. The membrane is then placed between two glass plates where a high voltage is applied for approximately five minutes. The voltage is then removed and the membrane is taken to be washed briefly in water and dyed with Reactive brown dye in water[4]. This entire process is short and is completed in about 40 minutes compared to the 2-D PAGE method which takes one to two days.

Manual results were first observed to find optimal conditions for protein separation. A high speed camera recorded the spotting process observing both the hand movement of the experimenter and also the spotting of the sample on the surface. It is from this observation, one is able to create a device in which optimal parameters are made repeatable. Figure 2 illustrates the results obtained for one manual experiment in which a mixture of six proteins was completely separated according to their respective isoelectric points. Positions 1-10 illustrate the protein separation that has occurred. Polarity is shown and each protein's corresponding isoelectric point is shown. Position 8 is the origin of spotting. The pH of the buffer solution is set at 5. Results shown in Figure 2 illustrate complete protein separation for this novel electrophoresis method.

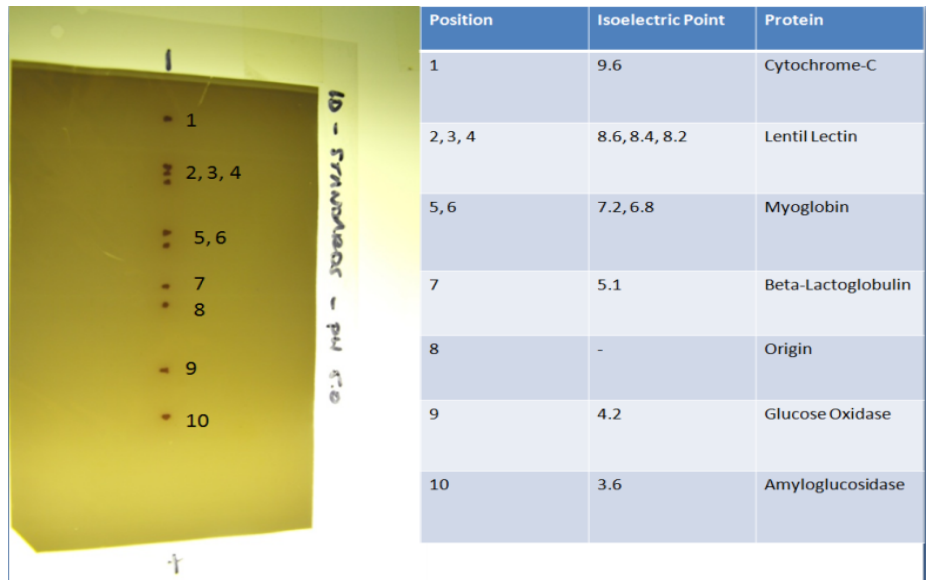

Figure 2. Manual Protein Spotting for a Six Protein Mixture

| Position | Isoelectric Point | Protein |
|---|---|---|
| 1 | 9.6 | Cytochrome-C |
| 2, 3, 4 | 8.6, 8.4, 8.2 | Lentil Lectin |
| 5, 6 | 7.2, 6.8 | Myoglobin |
| 7 | 5.1 | Beta-Lactoglobulin |
| 8 | - | Origin |
| 9 | 4.2 | Glucose Oxidase |
| 10 | 3.6 | Amyloglucosidase |

The automated methods proposed use the parameters from the manual spotting techniques known to be ideal. What is observed from these experiments is that proteins tend to concentrate in a donut like shape along the origin. Some binding is believed to occur because of this circular concentration which must be avoided to produce good results.

Our initial results indicate that some binding of the proteins occurs along the membrane surface using the automated methods. What is known from our experiments is that the reduced contact area and force application is most critical in producing good results.

Protein separation using this electrophoresis method is highly dependent on the experimenter. The application of force and the tip to membrane distance are believed to be the most critical parameters for producing good results. Extrapolating such ideal parameters is difficult to integrate in an automated device, as it is difficult to duplicate a human experimenter. The two automated methods proposed demonstrate that protein separation is possible when all known ideal parameters are taken into account. In experimenting with different membrane types in which pore size, cavity structure, and hydrophobicity are being reviewed to illustrate why manual results are so user dependent. It is because of this dependence, that automation is the only possible way of avoiding such a problem. It is our hope to improve the automated process such that the manual spotting is no longer necessary. It would result in an efficient experimental process that has merit in this industry.

## Acknowledgements

A special thanks to The Nanotechnology Institute, Philadelphia, PA for their financial support.